\documentclass[epj]{svjour}
\usepackage{graphics}
\usepackage{amssymb}
\usepackage{amsmath}
\usepackage{latexsym}
\usepackage{epsfig}
\usepackage{latexsym}
\usepackage{dcolumn}
\usepackage{bm}
\usepackage{float}
\usepackage{hyperref}
\usepackage{verbatim}
\usepackage{color}
\newcommand{\ct}{\cite}

\newcommand{\bi}{\bibitem}
\newcommand{\ba}{\begin{eqnarray}}
\newcommand{\ea}{\end{eqnarray}}

\newcommand{\non}{\nonumber}

\newcommand{\de}{\delta}

\def \beq{\begin{equation}}
\def \eeq{\end{equation}}
\def \bea{\begin{eqnarray}}
\def \eea{\end{eqnarray}}   
\def \ham{\mathcal{H}}        

\begin{document}

\title{Fidelity, Rosen-Zener Dynamics, Entropy and Decoherence in one dimensional hard-core bosonic systems}
\author{Sthitadhi Roy\inst{1}, Tanay Nag\inst{1} \and Amit Dutta\inst{1}
\thanks{\emph{Present address:} }%
}                     
\authorrunning{Roy, Nag and Dutta}
\titlerunning{Fidelity, RZ dynamics, Decoherence and Entropy in 1D Hard-Core Bosonic Systems}
\offprints{}          
\institute{Department of Physics, Indian Institute of Technology, Kanpur 208016, India.}
\date{Received: date / Revised version: date}
%
\abstract{
We study the non-equilibrium dynamics  of a one-dimensional system of hard core bosons (HCBs) in the presence of an onsite potential (with an  alternating sign between the odd and even sites) which shows a quantum phase transition (QPT) from the superfluid (SF) phase to the so-called "Mott Insulator" (MI) phase.   The ground state quantum fidelity shows a sharp dip at the quantum critical point (QCP) while  the fidelity susceptibility shows a divergence right there with its scaling  given in terms of the correlation length exponent of the QPT. We then study the evolution of this bosonic system following a quench in which the magnitude of the alternating potential is changed starting from zero (the SF phase) to a non-zero value (the MI phase) according to a half Rosen Zener (HRZ) scheme  or brought back to the initial value following a full Rosen Zener (FRZ) scheme. The  local von Neumann entropy density is calculated   
in the final MI phase  (following the HRZ quench) and is found to be  less than the equilibrium value  ($\log 2$) due to the defects generated in the final state as a result of the
quenching  starting from the QCP of the system. We also briefly dwell on the FRZ quenching scheme in which the system is finally in  the SF phase through the
intermediate MI phase and  calculate the reduction in the supercurrent and the  non-zero value of the residual local entropy density in the final state. Finally, the loss of coherence of a qubit (globally and weekly coupled to the HCB system) which is initially in a pure state is investigated by calculating the time-dependence of the decoherence factor when the HCB chain evolves under a HRZ scheme starting from the SF phase. This result is compared with that of  the sudden quench limit of the half Rosen-Zener scheme  where an exact analytical form of the  decoherence factor can be derived.
} 
\maketitle


\section{Introduction}\label{sec:intro}

Recent advancements in experiments on ultracold atoms trapped in optical lattices have facilitated the realization of ultracold vapors of bosonic atoms, and hence have opened up  new directions towards the experimental studies of low dimensional bosonic systems \cite{greiner_prl,bloch08}. For example,
following the pioneering experiments indicating  a superfluid (SF) to a Mott insulator (MI)  transition in optical lattices  in three-dimension \ct{greiner02}(and also in one dimension \ct{stoferle04}) and the corresponding study on the non-equilibrium dynamics \ct{sadler06},  there is an upsurge in the studies of quantum phase transitions (QPTs) \ct{sachdev99,chakrabarti96,sondhi97,continentino,vojta03} and dynamics  of trapped atoms in optical lattices. 
More interestingly, two dimensional  optical lattices have made the quasi one dimensional regime experimentally accessible \cite{greiner_prl,moritz}
 by keeping the transverse potentials much higher than the longitudinal potential. By appropriately  tuning the longitudinal potential, different limits of the bosonic Hubbard model have been realized. 
One of such  limits happens to be the hard-core boson (HCB) limit (or the Tonks-Girardeu \ct{tonks36,lenard66} limit), where two bosons
can not occupy the same site;  this  limit has also been achieved in an
optical lattice \cite{paredes,kinoshita}. 
These experiments have paved the way for a plethora of theoretical studies in low-dimensional bosonic
systems  \ct{polkovnikov06,cazalilla11} especially from the viewpoint of the SF to the MI transition \ct{altman02,fischer06} and related non-equilibrium dynamics \ct{sengupta04,tuchman06}. The HCB systems have turned out to be very advantageous in this
context \ct{rigol04,rousseau06,klich07}.

In parallel, there have been numerous studies which attempt to bridge a connection between QPTs\ct{sachdev99,chakrabarti96,continentino,sondhi97,vojta03} and
quantum information theoretic measures like concurrence \ct{osterloh02,amico08}, quantum fidelity\ct{zanardi06,gu10,venuti07,you07,zhou08,zhao09,gritsev09,schwandt09,albuquerque10,rams11}, quantum discord \ct{dillenschneider08}, entanglement entropy \ct{vidal03,kitaev061} etc.. These measures enable us to detect a QCP and they also show distinctive scaling relations close to it characterized by some of the associated critical exponents. Similarly, the decoherence (or loss of phase information) \ct{zurek03} of a qubit coupled to a quantum critical
system  is also being investigated\ct{rossini07,quan06}.

  The scaling of the density of defects (or heat) produced following a slow \ct{zurek05,polkovnikov05}
or rapid quenching \ct{grandi10} across (or starting from) a QCP has also attracted attention of the scientists.  Defects generated in the final state of the quantum system due to the quenching through a QCP in turn  lead to non-zero quantum correlations (for example, non-zero local
entropy density \ct{cherng06,mukherjee07}, concurrence \ct{sengupta09}, quantum discord \ct{nag11}, etc.)  in the final state  which are otherwise absent in the defect free final state. These  information theoretic measures have also been found  to satisfy scaling relations identical to that of the defect density in some cases.     For recent reviews, see [\cite{dutta10,polkovnikov11a,dziarmaga10}].

In this paper, we study the dynamics of a one-dimensional lattice of HCBs at half-filling in which Bosons are subjected to an
onsite potential. The model has 
a SF long-range order which persists up to a threshold value of the onsite potential at which 
there is a QPT from  the SF to the MI phase which is a chemical potential driven phase transition. Beyond the finite threshold value of the onsite potential (at which a gap opens up in the spectrum) the system becomes an insulator due to correlation effects and we have a Mott insulator in the true sense of the term. We are however interested in the case where the onsite potential is site-dependent (rather, alternates in sign on the even and odd sites); under this condition the SF long-range
order is destroyed as soon as the potential is switched on. We put a word of caution here; in our case the site-dependent onsite potential breaks the translation symmetry of the system and any non-zero value of this potential opens up a gap in the spectrum. Though it is not a MI in its true sense we continue to call it so as has been done in literature \ct{klich07}. We note that this model has been studied under a (HRZ) quenching scheme  \cite{rosen32,robiscoe78} in which the magnitude of the alternating onsite potential is quenched from zero to a non-zero value and the residual supercurrent in the MI phase has been estimated \ct{klich07}.

The motivation of this work is the following: although there has been a series of studies of quantum critical dynamics which involve Landau-Zener tunneling \ct{landau} (for many examples, see [\cite{dutta10,polkovnikov11a,dziarmaga10}]), the Rosen-Zener (RZ) tunneling (for  which the non-adiabatic excitation probability can also be exactly calculated) has received relatively less
attention. We use the integrability of the one-dimensional HCB system in an alternating potential along with  the exact analytical results for the  HRZ quenching to investigate the generation of local entropy in the HCB system in its final MI  state following the quench and also the reduction in the supercurrent and residual local entropy in the SF phase following the FRZ quench.
We also calculate the decoherence of a qubit connected
to the HCB system following a HRZ quenching of the magnitude of the onsite potential.  Given the current interest in QPTs, dynamics and quantum information as discussed above, these results are expected to 
be useful  both from experimental and theoretical viewpoints. 

The paper is organized in the following way: in Sec. \ref{sec:model},  we describe the QPT in the HCB chain in an alternating potential by analyzing
the energy spectrum of the Hamiltonian; any non-zero value of the  alternating potential  leads to an energy gap in an otherwise
gapless spectrum so that the system is in the MI phase. In Sec. \ref{sec:fidelity}, we show  how this QPT 
can be detected and characterized  by investigating the ground state fidelity and fidelity susceptibility.

The dynamics of the HCB chain
is studied in Sec. \ref{sec:RZ}. 
in Sec.\ref{subsec:hrz}, we investigate the single site (local) von Neumann entropy density  in the final MI phase following the  HRZ quenching for the HCB system.
We note that
the local entropy density is zero in the SF phase and is equal to $\log 2$   in the MI phase because of its bipartite structure.  We, however, find that  the value of this  entropy in the final MI phase reached after the quenching is less than $\log  2$ by an amount which depends on the parameters of the HRZ quenching. This deviation is due to the fact that the system is quenched out of the SF phase (which is also a gapless QCP) at a finite rate which leads to the defects resulting in  a surviving supercurrent and reduced local entropy density in the
 final MI phase. 
In Sec.\ref{subsec:frz}, we study  the HCB chain under the
full Rosen Zener (FRZ) quenching scheme    in which the system is finally brought back to the SF phase
through the intermediate MI phase and  
 the surviving supercurrent and the residual local entropy  density are  calculated.

Finally in Sec.\ref{sec:deco} a qubit (or a central spin-1/2) is globally coupled to the HCB chain.
Our focus  is limited to the case when the coupling between the qubit and the HCB chain, which in fact  plays the role of an environment to which the qubit is coupled, is very weak. We study the decoherence of the qubit by
calculating  the decoherence factor in the final state when the onsite potential is changed from zero (the SF phase) to a finite value (the MI phase) following a HRZ quenching scheme in Sec.\ref{sec:decohrz}.
An exact expression of the decoherence factor of the qubit is derived analytically  in  the sudden quench limit in Sec.\ref{subsec:decosq}  where the alternating potential is instantaneously switched on  and  the results  are compared to those of the previous case.

\section{The Model}\label{sec:model}
We consider  the Lattice-Tonks-Girardeu gas (hard-core) limit of the one-dimensional Bosonic Hubbard model \cite{cazalilla11}  given by the Hamiltonian 
\beq
\ham = -w\sum_l(b_l^{\dagger}b_{l+1} + \text{h.c}) + V\sum_l(-1)^l b_l^{\dagger}b_l,
\label{eq_ham1}
\eeq 
where $w$ is the hopping amplitude, $V$ is the onsite potential; $b_l$ and $b_l^{\dagger}$ are the bosonic annihilation and creation operators at the $l^{th}$ site of the lattice, respectively.  These bosonic operators satisfy the canonical commutation relation 
$[b^{\dagger}_l,b_m] = \delta_{lm}$;
additionally, the hard core condition demands,  $(b_l)^2 = 0 = (b_l^{\dagger})^2$. The Hamiltonian (\ref{eq_ham1}) undergoes a QPT from the gapless SF phase to the gapped MI phase for any non-zero value of the alternating potential $V$ as shown below.

\begin{figure}
 \begin{center}
  \includegraphics[width=0.8\columnwidth]{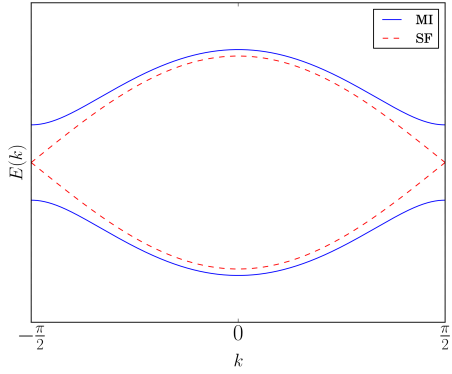}
 \end{center}
\caption{(Color online) The energy spectrum (\ref{eq_ek}) for the model (\ref{eq_ham1}) in the reduced Brillouin zone. In the SF phase ($V=0$), the spectrum is gapless at $k=\pi/2$ (red dashed). 
A non-zero $V$generates a gap in the excitation spectrum at the critical modes are at $k=\pm \pi/2$.}
\label{fig_ek}
\end{figure}

This Hamiltonian can be exactly solved using Jordan-Wigner (JW) transformations\cite{lieb61} given by 
\beq
b_l^{\dagger} =\left[\prod_{m<l}\text{exp}\left(a_m^{\dagger}a_m\right)\right]a_l^{\dagger},
\label{eq_jw}
\eeq where $a_l^{\dagger}$ and $a_l$ are the JW fermionic operators satisfying the fermion anti-commutation relations
$\{a^{\dagger}_l,a_m\} = \delta_{lm},  \{a_l,a_m\}=0$.
Using JW transformation followed by the  Fourier transformation, the energy spectrum of Hamiltonian (\ref{eq_ham1}) can be exactly obtained.  
In terms of JW fermions, the Hamiltonian can be re-written as
$\ham = \ham_0 +\ham_d $, where,
\beq 
\begin{split}
&\ham_0 = -\sum_{|k|<\pi/2} 2w\cos k(a_k^{\dagger}a_k - a_{k+\pi}^{\dagger}a_{k+\pi}),\\ 
&\ham_d = \sum_{|k|<\pi/2} V(a_{k+\pi}^{\dagger}a_k + a_{k}^{\dagger}a_{k+\pi}). \label{eq_ham2}
\end{split}
\eeq
Evidently,  the mode with wave vector $k$ couples to the  $(k+\pi)-$ mode, one can rewrite the Hamiltonian in the reduced $2 \times 2$ form,
\beq
\ham= \otimes \sum_{|k|<\pi/2}\ham_k,
\label{eq_ham3}
\eeq 
with
\beq
\ham_k = \begin{pmatrix}
             2w\cos k & -V\\
             -V & -2w\cos k
            \end{pmatrix}
\label{eq_hkmatrix},
\eeq
and the energy spectrum (see Fig.(\ref{fig_ek})) is given by
\beq
E_k = \sqrt{4w^2\cos^2 k + V^2}.
\label{eq_ek}
\eeq
We note that the spectrum (\ref{eq_ek}) is gapped even for an infinitesimal alternating potential implying that the system is in the MI phase for any 
$V \neq 0$. On the other hand, for $V=0$, the spectrum is gapless for the critical mode $k=\pi/2$, and the HCB chain is in the SF phase. It should be noted that the critical modes at $k=\pm\pi/2$ are also the Fermi levels since we are working at half-filling. From the spectrum, we find that the QPT at $V=0$ is characterized by the correlation length exponents $\nu=1$ and  the dynamical exponent $z=1$.

\section{Fidelity and Fidelity Susceptibility}\label{sec:fidelity}

One of the most widely used quantum information theoretic measure for detecting and characterizing quantum phase transitions is ground state  quantum fidelity \cite{gu10,venuti07,you07} which is the magnitude of the overlap of the two ground states of a
quantum many body system belonging to different values of a parameter of the Hamiltonian. Referring to the Hamiltonian (\ref{eq_ham1}), we can define quantum fidelity $F(V,V+\delta)$ between two ground states with the alternating potentials  $V$ and $V+\delta$, respectively, given by 
\ba
 F = |\left<\psi_{0}(V)|\psi_{0}(V+\delta)\right>| = 1 - \frac{\delta^{2}}{2}L^{d}\chi_{F}+\cdots,
 \label{eq_fid}
 \ea
 where we have assumed a small system size ($L$) and also $\delta \to 0$ limit, which allow us to truncate the above series at the $\delta^2$ order; in the present problem spatial dimensionality $d=1$.
The quantity $\chi_{F} = -(2/L^{d})\text{ln}(F)/\delta^{2}\vert_{\delta \rightarrow 0} $, called the fidelity susceptibility density \cite{venuti07,you07,zhou08,zhao09,gritsev09,schwandt09,albuquerque10,rams11}, is a measure of the rate of the change of the ground state wave function when the parameter $V$ is changed infinitesimally.
Usually quantum fidelity shows a sharp dip at a QCP where $\chi_F$ diverges with the system size; the universal scaling of $\chi_F$ is given in terms of some of the critical exponents associated with the QPT.

\begin{figure}
	\begin{center}
		\includegraphics[width=0.9\columnwidth]{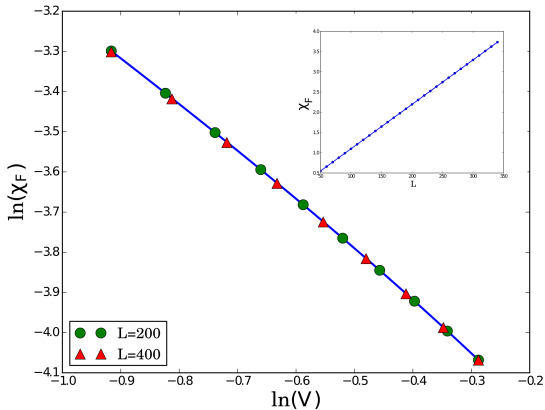}
 \end{center}
\caption{(Color online) Numercially obtained scaling of $\chi_F$: away from the QCP, $\chi_F \sim V^{-1}$,
while in the vicinity QCP, $\chi_F \sim L$ (see inset). These scaling relations are in agreement with the
theoretical prediction.}
\label{fig_fidsus_scaling}
	\end{figure}

To calculate $F$ and $\chi_F$ in the vicinity of the QPT of Hamiltonian (\ref{eq_ham1}), we use the reduced
two-level Hamiltonian (\ref{eq_hkmatrix}).
One can use Bogoliubov transformation to obtain the ground state wave function for a particular momentum mode  and a given potential $V$ in the form
 \beq
|\psi_0(k,V)\rangle = \cos(\theta_{k}(V))|k\rangle + \sin(\theta_{k}(V))|k+\pi\rangle
\label{eq_gs}
\eeq
where $\tan (2\theta_{k} (V)) = -{V}/{(2w\cos k)}$. An exact expression of quantum fidelity can be then obtained  using Eqs.~(\ref{eq_fid}) and (\ref{eq_gs}):
\begin{align}
F&= \prod_{k}|\cos(\theta_{k}(V)-\theta_{k}(V+\delta))|\non\\
&= \text{exp}\left[\frac{L}{2\pi}\int_{-\frac{\pi}{2}+\frac{\pi}{L}}^{\frac{\pi}{2}-\frac{\pi}{L}}dk\:\log(|\cos(\theta_{k}(V)-\theta_{k}(V+\delta)|)\right] \label{eq_fidk}.
\end{align}
\begin{figure}
	\begin{center}
		\includegraphics[width=0.9\columnwidth]{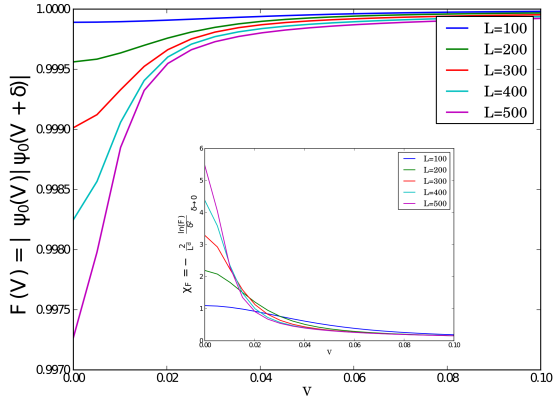}
 \end{center}
\caption{(Color online) Fidelity shows a clear dip at $V=0$. The inset shows that a peak occurs in the fidelity susceptibility at the critical point. Clearly, the value of fidelity drops from unity even away
from the QCP as the system size increases. }
\label{fig_fidelity}
	\end{figure}
  We also find that $\chi_F$ scales as $L$ near $V=0$ (SF phase) and as $V^{-1}$ deep inside MI phase (see
 Fig.(\ref{fig_fidsus_scaling})) . Expanding around the critical mode $k=\pi/2$,  one arrives at the simplified form

\begin{equation}
\begin{split}
F = \text{exp}\left[\right.\frac{\delta^{2}}{32\alpha V^2}\left\{\right. &\tan^{-1}(\alpha)-\tan^{-1}[\alpha(L-1)] \\
  &-\frac{\alpha}{1+\alpha^{2}} + \frac{\alpha(L-1)}{1+\alpha^{2}(L-1)^{2}}\}];\:\:\alpha = \frac{2w\pi}{VL}.
\end{split}
\label{eq_fidexp}
\end{equation}
The expansion around the critical mode is meaningful because the integrand in the argument of the exponential in Eq.(\ref{eq_fidk}) goes to zero near the critical modes. For modes  away from the critical mode,  the integrand is  highly negative and hence their contribution to fidelity is vanishingly
small for large $L$. As shown in Fig.(\ref{fig_fidelity}), the fidelity shows a  dip and the susceptibility shows a peak at the QCP, $V=0$. This is in congruence with the generic scaling\cite{gritsev09,schwandt09,albuquerque10,rams11},  $\chi_F\ \sim L^{2/\nu -d}$ near the QCP ($ L \ll V^{-\nu}$), and $\chi_F\sim V^{\nu d -2}$ away from the QCP ($ L \gg V^{-\nu}$),  with $\nu=d=1$.

\section{RZ quenching of the on-site potential}\label{sec:RZ}

In this section, we shall study the HCB model under the HRZ and FRZ quenching schemes and calculate the von Neumann entropy and the diagonal entropy following the HRZ quench and the supercurrent density and the von Neumann entropy following the FRZ quench.

\subsection{Von Neumann entropy and Diagonal entropy of the HCB chain following the HRZ quench}\label{subsec:hrz}

\begin{figure}[H]
	\begin{center}
		\includegraphics[width=0.9\columnwidth]{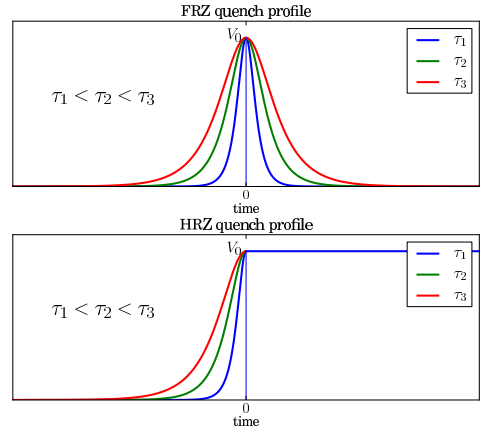}
 \end{center}
\caption{(Color online) The FRZ and HRZ quenching schemes for $V(t)$. We get the sudden quench limit by taking $\tau\rightarrow 0$ in the HRZ case.}
\label{fig_rz}
	\end{figure}

In this subsection, we shall employ the HRZ quenching scheme in which the alternating potential 
is changed from zero to a finite value $V_0$,  (see Fig.~(\ref{fig_rz})) in  a non-linear fashion  given by \cite{robiscoe78,klich07}
\beq
V(t) = \left\{\begin{matrix}
               V_0\left(\text{sech}\left(\frac{\pi t}{\tau}\right)\right)\; \; & t<0\\
               V_0 &t \ge 0.
              \end{matrix}\right.\label{eq_rz}
\eeq              
This implies that the system is quenched from the SF phase ($t \to -\infty$) to the MI phase ($t=0$). 

In order to calculate the time evolution of $|\psi(t)\rangle$  at a given instant $t$, let us consider a generic 
state for a given momentum mode: $|\psi_k(t)\rangle = s (t)|k\rangle + p (t)|k+\pi\rangle$. Using Schr$\ddot{o}$dinger equations $i\partial|\psi _k(t)\rangle/\partial t = \ham_k |\psi _k(t)\rangle$, it can be shown that time evolution of the probability amplitudes $s (t)$ and $p (t)$ are dictated by the equations,  \cite{klich07,rosen32}
\beq
\begin{split}
 &i\dot{s} (t) = s (t)2w\cos(k) + p (t)V (t),\\
 &i\dot{p} (t) = -p (t)2w\cos(k) + s (t)V (t).
\end{split}
\label{eq_diffsp}
\eeq

Using transformations  $S (t) = \exp(2iwt\cos{k})s (t)$, $P (t)=\exp(-2iwt\cos{k})p (t)$, we get
\beq
\begin{split}
\ddot{S } =& -\left(\frac{V_0 }{\text{cosh}\left(\frac{\pi t}{\tau }\right)}\right)^2S \\ & + \left[4iw\cos(k)-\frac{\pi}{\tau }\text{tanh}\left(\frac{\pi t}{\tau }\right)\right]\dot{S },
\end{split}
\label{eq_diffS}
\eeq
which  can be reduced to a hypergeometric form with the initial conditions, $|S (-\infty)|=1$ and $|P (-\infty)|=0$. Expanding near the critical mode ($k  = \pi/2$), one eventually finds the solution at $t=0$ of the form:
\beq
\begin{split}
 &p (0) = -i\sin\left(\frac{V_0 \tau }{2}\right),\\ 
 &s (0) = \cos\left(\frac{V_0 \tau }{2}\right).
 \label{eq_s0p0}
\end{split}
\eeq
Exploiting the continuity condition of the wave function at $t=0$, let us write the generic wave function for $t > 0$ in the form
\beq
\begin{split}
\vert\psi _k(t)\rangle &= c_g  \vert g  (t) \rangle + c_e  \vert e  (t)\rangle\\
&=c_g  e^{iE_k  t}\vert g  (0) \rangle + c_e  e^{-iE_k  t}\vert e  (0)\rangle,\\
\label{eq_ge}
\end{split}
\eeq
where $|g  \rangle$ and $|e  \rangle$ are the ground state and excited state wave functions (with energies $-E_{k}  $ and $E_{k}  $) with probability amplitudes $c_g  $ and $c_e  $, respectively. Expressing Eq.~(\ref{eq_ge}) in terms of momentum modes $|k\rangle$ and $|k+\pi\rangle$ and using Bogoliubov transformation, we get  
\beq
\begin{split}
\vert \psi_k  (t)\rangle =& (s  (0)A_k(t)   + p  (0)B_k(t)  )\vert k \rangle + \\
                                         & (s  (0)B_k(t)   + p  (0)A^{*}_k(t))\vert k+\pi \rangle, 
\label{eq_finalstate}
\end{split}
\eeq
where $A_k (t) = \cos(E_k  t) +i \cos(2\theta _k)\sin(E_k  t)$, $B_k (t) =i\sin(2\theta _k)\sin(E_k  t)$
and $E_k  = \sqrt{4w^2\cos^2k + V_0^2}$.

Using the wave function following the quench at an instant $t$ given in Eq.~(\ref{eq_finalstate}), we are now in a position to calculate the single-site von Neumann entropy given by $ -\text{Tr}\; \rho\log(\rho)$ where $\rho$ is the density matrix constructed from $\vert \psi_k  (t)\rangle$. Ideally in the MI phase, the local von Neumann entropy density  $s=\log 2$. (The MI phase is in a pure state and hence the global entropy is zero. However, the (single site) local entropy obtained by integrating over the momentum modes is non-zero because of the bipartite structure of the MI phase. Interpreting in terms of the spin variables,  when observed locally upon ``coarse-graining" in momentum \ct{cherng06}  both the spin states appear with an equal probability ($=1/2$) which makes the  entropy density $\log 2$ ).

In the present context, however,  the MI phase is reached through a non-equilibrium variation of the alternating on-site potential  starting from the SF phase at a finite rate and hence the entropy density in the MI phase gets reduced.
To calculate it, we decompose  the density matrix in a direct product form,  $ \rho = \bigotimes\prod_k \rho_k$, where $\rho_k$ is the reduced density matrix for the $k$-th  mode. Consequently, the entropy density turns to be 
 \begin{equation}
  s = -\frac{1}{\pi}\int_{-\pi/2}^{\pi/2}dk\;   \rm{Tr}~\rho_k\log(\rho_k)
  \label{eq_vnedef}
 \end{equation} 
 To calculate $\rho_k$ following the HRZ, we use Eq.(\ref{eq_finalstate}); we are interested in the long-time average of $s$ and since the integrals over $k$ and $t$ commute we can take the long-time average of the terms of $\rho_k$ itself before doing the integral over $k$. Taking the long time average of the terms of $\rho_k$, we find  

\begin{equation}
 \rho_k = 
 \begin{pmatrix}
  \frac{\left(1+\cos^22\theta_k\cos(V_0\tau)\right)}{2} & \frac{\cos2\theta_k\sin2\theta_k\cos(V_0\tau)}{2}\\
  &\\
  \frac{\cos2\theta_k\sin2\theta_k\cos(V_0\tau)}{2} & \frac{\left(1-\cos^22\theta_k\cos(V_0\tau)\right)}{2}
  \label{eq_dmat}
 \end{pmatrix}.
\end{equation}

\begin{figure}
 \begin{center}
  \includegraphics[width=0.9\columnwidth]{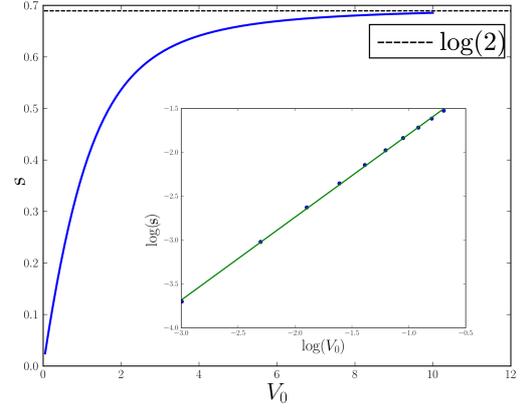}
 \end{center}
\caption{Variation of $s$ with $V_0$ in the MI phase (with $w=1$). The inset justifies that $s$ scales linearly with $V_0$ for small $V_0$.
 } 
\label{fig_vne_hq_V}
\end{figure}

\begin{figure}
 \begin{center}
  \includegraphics[width=0.9\columnwidth]{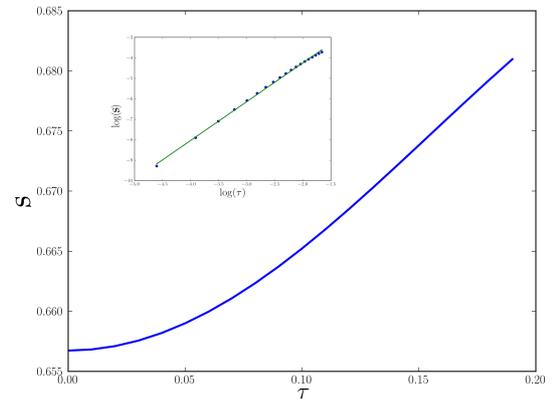}
 \end{center}
\caption{Variation of $s$ with $\tau$ in the MI phase with $w=1$. The inset shows that $\log s$ scales lineally with  $\log \tau$ with a slope
=2.}
\label{fig_vne_hq_tau}
\end{figure}

Diagonalizing the density matrix, the entropy density can be expressed in terms of the eigenvalues $\lambda_{\pm,k}$ as
 \begin{equation}
  s = -\frac{1}{\pi}\int_{-\pi/2}^{\pi/2}dk\;  \lambda_{+,k}\log(\lambda_{+,k})+\lambda_{-,k}\log(\lambda_{-,k}),
\end{equation}
where$\lambda_{\pm,k} =  \frac{1}{2}\left(1\pm\cos2\theta_k\cos(V_0\tau)\right)$.
The von Neumann entropy density  $s$ increases linearly with $V_0$ for  $V_0 < 2w$, and saturates to the maximum value of $\log 2$ for higher values of $V_0$ (see Fig.~(\ref{fig_vne_hq_V})). On the other hand,  $s$ is found to scale  quadratically with $\tau$ (see Fig.~(\ref{fig_vne_hq_tau})). As mentioned
already that for a HRZ quenching, the parameters $V_0$ and $\tau$ are not on an identical footing which is also
reflected in the scaling of $s$. 


\begin{figure}
\begin{center}
\includegraphics[width=0.8\columnwidth]{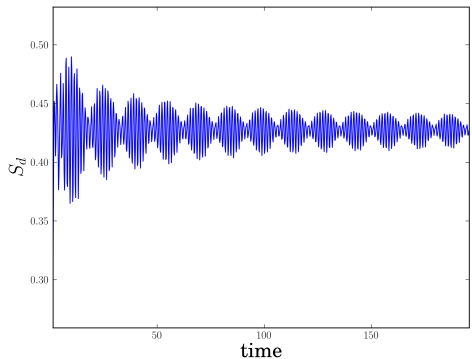}
\end{center}
\caption{The diagonal entropy density $S_d$ is plotted against time shows an oscillatory  interference pattern similar to that found in the surviving supercurrent  in the MI phase after a similar  as reported in  \cite{klich07}.}
\label{fig_diagent}
\end{figure}

One can also calculate the diagonal entropy \cite{polkovnikov11b} defined as $S_d(t) = \sum_n \rho_{nn}(t)\log\rho_{nn}(t)$ where $\rho_{nn}(t)$ are the diagonal terms of the density matrix obtained from Eq.~(\ref{eq_finalstate}) (without any time averaging).  The diagonal entropy  $S_d(t)$ shows an oscillatory behavior (see Fig.~(\ref{fig_diagent})) similar to the supercurrent in the MI phase following a similar quench \cite{klich07}. The scaling of the diagonal entropy $S_d$ with $V_0$ and $\tau$ is same as compared to the scaling of von Neumann  entropy density $s$ in both the region of $V_0$.

\subsection{Current and von Neumann entropy studies after a FRZ quench}\label{subsec:frz}
In this subsection, we shall estimate the supercurrent  and von Neumann entropy following a FRZ quench of the HCB chain (without the qubit) using the time-evolution of the potential  given by the following form:
\beq
V(t) = V_0\: \text{sech}\left(\frac{\pi t}{\tau}\right), \: -\infty<t<+\infty;
\label{eq_fullrz}
\eeq 
the system is  initially ($t \to -\infty$) in the SF phase and finally brought back to the SF phase (as $t \to \infty$) through the  intermediate MI phase. We study the time evolution of the system after the quenching process gets  over (i.e., in the final SF phase). In the SF phase the reduced Hamiltonian is diagonal  in the basis $\vert k \rangle$ and $\vert k+\pi\rangle$   (with $\vert k \rangle$  ($\vert k+\pi\rangle$) being the ground state (excited state)).  The wave-function of  the HCB system immediately after the FRZ quench (which we set as $t=0$) can be written as a linear combination of these basis states, 
\beq
\vert\psi(t=0)\rangle = \sqrt{1-P_k}\vert k \rangle + \sqrt{P_k}\vert k+\pi\rangle,
\label{eq_frzini}
\eeq
where $
P_k$ is the RZ non-adiabatic transition formula \ct{rosen32}
\begin{equation}
 P_k = \sin^2\left(\frac{V_0\tau}{2}\right)\text{sech}^2[2\tau w\cos{k}].
 \label{eq_pk}\end{equation}

The time-evolved wave-fuction at some later time $t$ can readily be written as
\beq
\vert\psi(t)\rangle = \sqrt{1-P_k}e^{-iE_kt}\vert k \rangle + \sqrt{P_k}e^{iE_kt}\vert k+\pi\rangle,
\label{eq_rzwv}
\eeq
where $E_k=-2w\cos k$ in the SF phase. In order to calculate supercurrent one has to apply a boost to the Hamiltonian which takes the form $-w\sum_l (e^{-i\nu} b_l^{\dagger}b_{l+1} + \text{h.c})$ \cite{klich07}; consequently, the momentum value gets shifted from $k$ to $k+\nu$. Expressing the super-current operator, $\hat{j}(t) = \frac{iw}{L}\sum_l(e^{-i\nu}b^{\dagger}_{l+1}b_l - \text{h.c})$, in the Fourier space, one can find its expectation value with respect to the boosted counterpart of the state (\ref{eq_rzwv})
\ba
j(t) &=& \frac{w}{\pi}\int_{-\pi/2}^{\pi/2}dk\;\;\;\; \sin(k+\nu)(1-2P_{k+\nu}), \nonumber\\
&=& \frac{2w\sin{\nu}}{\pi}\left[1-2\left(V_0\tau\right)^2\right]
\label{eq_frzjt}
\ea

\begin{figure}
\begin{center}
\includegraphics[width=0.8\columnwidth]{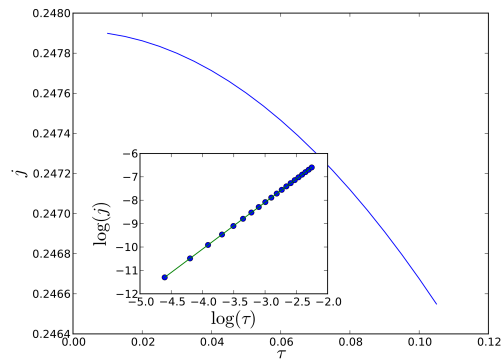}
\end{center}
\caption{(color online) The figure shows the variation supercurrent $j$ against $\tau$. The inset shows the plot of $\log(j)$ versus $\log(\tau)$, which is a straight line with slope 2}
\label{full_current}
\end{figure}
The above result leads the following interesting observations: in the limit of small $\nu$, $\sin(\nu)\sim\nu$, $j\sim\nu$ which is identical to the $\nu$ dependence of the supercurrent  in the initial SF phase. Secondly, the supercurrent becomes independent of time after the FRZ quench. This is due to the fact that at the final time HCB system reaches its eigen states. Thirdly, because of the passage through the MI phase starting from a QCP, the current in the final state is reduced from its initial value (at $t \to -\infty$), $(2w \sin\nu/)\pi$, by
a factor $V_0^2\tau^2$. It is also  to be noted that the the correction term of the supercurrent is  a function of the
combination $V_0 \tau$ implying that $V_0$ and $\tau$ are on the same footing for the FRZ quenching. This result is numerically verified as shown in Fig.~(\ref{full_current}).

\begin{figure}
 \begin{center}
  \includegraphics[width=0.9\columnwidth]{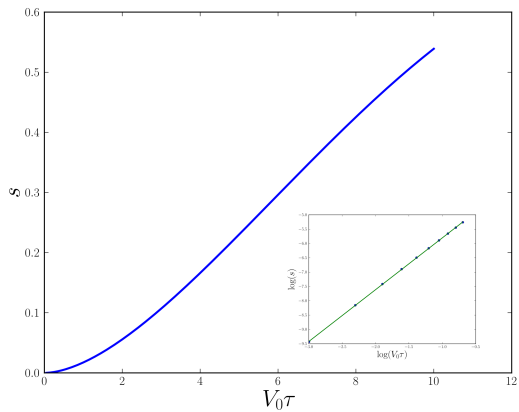}
 \end{center}
\caption{(color online)Variation of $s$ with $V_0\tau$ in the SF phase. The inset shows that $s\sim V_0^2\tau^2$ for small $V_0$ and $\tau$.} 
\label{fig_vne_fq}
\end{figure}

 In a similar spirit, one can calculate the residual von Neumann entropy density in the final SF phase.  The rapidly
oscillating off-diagonal terms of the reduced density matrix constructed from the wave function given in Eq.~(\ref{eq_rzwv}), vanish over long time averaging \ct{cherng06,mukherjee07} so that the decohered reduced
density matrix has a diagonal form.  Calculating the local entropy density using this decohered reduced denstity matrix, one can  show that $s\sim V_0^2\tau^2$ (see Fig.(\ref{fig_vne_fq})). One interesting point should be highlighted here: the quenching through the MI phase generates defects in the SF phase which result in a
 reduction of the supercurrent  
and a  non-zero value of $s$  both scaling as $V_o^2 \tau^2$.

\section{Decoherence and HRZ quenching}\label{sec:deco}
 
 \subsection{Decoherence following a HRZ quencht}\label{sec:decohrz}
 In this section, we shall  explore the decoherence of a qubit coupled to the environment, chosen to be the HCB chain (\ref{eq_ham1})),  
 which is driven
 following the HRZ quenching scheme.
We assume a global coupling between  the qubit and  all the bosons of the model~(\ref{eq_ham1})  with the coupling Hamiltonian given by
\beq
\ham_{\mathcal{SE}} = -\delta\sum_lb_l^{\dagger}b_l\sigma^z_{\mathcal{S}},
\label{eq_ham_se}
\eeq
where $\sigma^z_S$ represents the qubit, $b_l^{\dagger}b_l$ is the number density of the environmental HCB chain at site $l$; $\delta$ is the coupling parameter between system and environment. 
(The form of the coupling Hamiltonian (\ref{eq_ham_se}) can be interpreted in the following way: the HCB chain can be recast to
a transverse XY spin chain in a transverse field in the $z$-direction; the $z$ component of the spin at the site $l$ is coupled to the
$z$-component of the central qubit.)
In subsequent sections, We shall work in the limit of a weak coupling between the central qubit and HCB system (i.e., $\delta \to 0$).

Due to the coupling to the central qubit,  the time evolution of the environmental bosonic chain is split into two channels, corresponding to the $|\left\uparrow\right>( \equiv +1)$ and  $|\left\downarrow\right>(\equiv -1)$ state of the the qubit. 
Using Eq.~(\ref{eq_hkmatrix}), we find that the reduced Hamiltonians of  the HCB system for  these two channels,
 denoted by $\ham_k^+$ and $\ham_k^-$, respectively, are given by
\beq
\ham_k^\pm = \begin{pmatrix}
             2w\cos{k} & -(V\pm\delta)\\
             -(V\pm\delta) & -2w\cos{k}
            \end{pmatrix}.
\label{eq_hkse}
\eeq
We shall denote the corresponding  time-evolved states of the environmental Hamiltonian corresponding to these two branches as $\vert \psi^+(t)\rangle$ and $\vert \psi^-(t)\rangle$, respectively.

\begin{figure}
	\begin{center}
		\includegraphics[width=0.9\columnwidth]{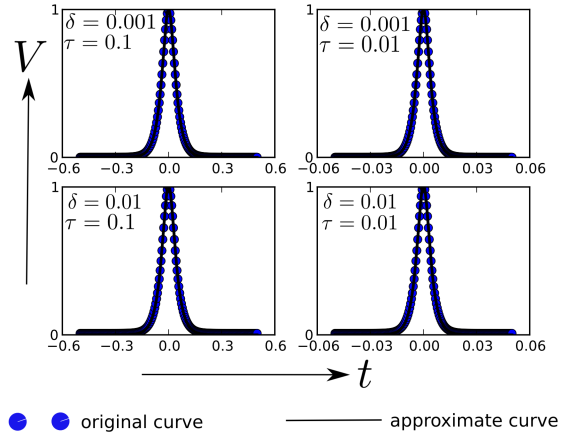}
 \end{center}
\caption{For $\delta \to 0$ and small $\tau$, we verify the approximation made in Eq.(\ref{eq_rzpm}) by plotting the exact expression and the approximate expression for all time. They are in a very good agreement with each other.}
\label{fig_deltafit}
	\end{figure}

One can show that in the limit $\delta \to 0$, the off-diagonal terms of the Hamiltonian(\ref{eq_hkse}) can be written as
\beq
V(t)\pm\delta=V^\pm(t) = (V_0\pm\delta)\text{sech}\frac{\pi t}{\tau^\pm},
\label{eq_rzpm}
\eeq
where  $\tau^\pm = \tau \pm \alpha\delta\tau$, with $\alpha$ being a constant of the order of unity. It can be shown numerically that $\alpha$ is not a function of time as well (see Fig.\ref{fig_deltafit}). It is to be noted that we have made a a small $\tau$ approximation; we intend to study the dynamics close to the sudden quench limit.

When compared with the
RZ form Eq.~(\ref{eq_rz}), this approximation implies the
following: in the limit of a very weak coupling between the qubit and the environment, the evolution of the two channels
can be viewed as two independent HRZ quenches with final potentials and quenching parameters $[(V_0+\delta)$,
$\tau^{+}]$ and $[(V_0-\delta)$, $\tau^{-}]$, respectively. 

To study the decoherence of the qubit  coupled to the HCB chain following the HRZ quench ($t>0$), one investigates
the reduced density matrix of the  qubit. We assume that the qubit is initially in a pure state at $t \to -\infty$. The off-diagonal terms
of the reduced density matrix for $t>0$ incorporate the decoherence factor $D(t) = \vert\langle\psi^+(t)\vert\psi^-(t)\rangle\vert^2$,
which measures the decoherence of the qubit. A non-zero value (less than unity) of $D(t)$ implies that the qubit is in a mixed state and initial phase coherence is lost. Considering the two-level structure of the reduced Hamiltonian of the environmental HCB chain (see (\ref{eq_hkse})) we get,

\begin{equation} 
D(t) = \prod_k \vert D_k(t)\vert^2;~~~~ D_k(t) =  \vert\langle\psi_k^+(t)\vert\psi_k^-(t)\rangle\vert,
\end{equation}
which can be put in the form
\beq
 D(t) = \vert\langle\psi^+(t)\vert\psi^-(t)\rangle\vert^2 = {\exp}\left(\frac{L}{2\pi}\int_{k=-\pi/2}^{\pi/2}{\ln}(\vert D_k(t)\vert^2)dk\right). 
 \label{eq_deco1} 
 \eeq

To evaluate $D(t)$, we now use Eq.~(\ref{eq_finalstate}) and work in the limit $\delta \to 0$, when one can approximate
$\theta^{\pm}_k$ (defined after Eq.~(8) with $V \to V_0 \pm \delta$) as 
\beq
\theta^{\pm}_k = \theta_{k} +\delta\left. \frac{\partial \theta^{\pm}_k}{\partial \delta}\right|_{\delta=0},
\eeq
where
\beq
\left.\frac{\partial \theta^+_k}{\partial \delta}\right|_{\delta = 0} = -\left.\frac{\partial \theta^-_k}{\partial \delta}\right|_{\delta = 0} = \frac{-2w\cos k}{4w^2\cos^2 k + V_0^2}.
\label{eq_deriv}
\eeq
One can obtain using Eq.~(\ref{eq_deriv})
\beq
\begin{split}
 &\langle\psi_k^+(t)\vert\psi_k^-(t)\rangle = \cos(E_k^+t-E_k^-t)\cos(2\gamma)+\\
 &\sin(E_k^+t-E_k^-t)\sin(2\gamma)\left(\sin2\theta_{k}+2\cos2\theta_{k}\left.\delta\frac{\partial \theta^+_k}{\partial \delta}\right|_{\delta = 0}\right)-\\
 &i\sin(E_k^+t-E_k^-t)\cos(2\phi)\left(\cos2\theta_{k}+2\sin2\theta_{k}\left.\delta\frac{\partial \theta^-_k}{\partial \delta}\right|_{\delta = 0}\right),
\end{split}
\label{eq_psipluspsiminus}
\eeq
where $\gamma = \delta\tau(1+V_0\alpha)/2$ and $\phi = V_0\tau/2$.

The maximum  contribution to the  Eq.~(\ref{eq_psipluspsiminus}) comes from  the modes close to the critical mode $k =\pi/2$. We assume small $\tau$ ($\tau \ll 1$) so that  $\delta\tau  \to 0$ and  use the fact that $\ln(1-x) \sim -x$; the decoherence factor in the early time limit 
gets reduced to the form
\beq
\mbox{ln}(\vert D_k(t)\vert^2) =  - \delta^2t^2\frac{4V_0^2}{4w^2k'^2+V_0^2}(1+4\tau^2w^2k'^2),
\label{eq_logdeco}
\eeq
where $k'=\pi/2-k$. Using Eq.~(\ref{eq_logdeco}) and Eq.~(\ref{eq_deco1}) and retaining terms up  to the leading orders in $\delta$ and $\tau$, we get 

 \beq
 \mbox{ln}D(t) = \left[\frac{-L\delta^2t^2}{2\pi}\left\{\frac{2V_0}{w}\tan^{-1}\left(\frac{2\pi w}{V_0}\right)\left(1-V_0^2\tau^2\right) + 4\pi V_0^2\tau^2\right\}\right].
 \label{eq_decofinal}
 \eeq

\begin{figure}
	\begin{center}
		\includegraphics[width=0.9\columnwidth]{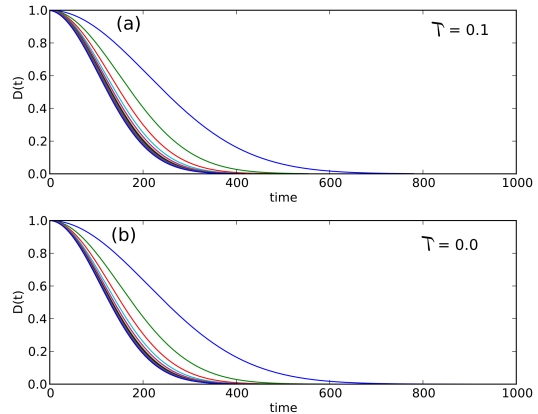}
 \end{center}
\caption{(Color online)(a) Gaussian fall of $D(t)$ following a HRZ quenchis shown with time. The different lines correspond to different values of $V_0$. Increasing $V_0$ makes the decoherence fall faster. The range of $V_0$ covered in the plot is from 1 (blue) to 50 (black). For $V_0>2\pi w$, $D(t)$ is approximately independent of $V_0$ which is refelcted by the bunching up of the curves for different $V_0$ for higher values. (b) The similar nature is observed for $D(t)$ in the sudden quench case ($\tau=0$).}
\label{fig_deco_t}
	\end{figure}

Eq.~(\ref{eq_decofinal}) is plotted in Fig.~(\ref{fig_deco_t}) which shows a gaussian fall in time of the decoherence factor in the early time limit; this Gaussian fall is the expected behavior in the vicinity of a QCP \ct{quan06}.

\begin{figure}
	\begin{center}
		\includegraphics[width=0.9\columnwidth]{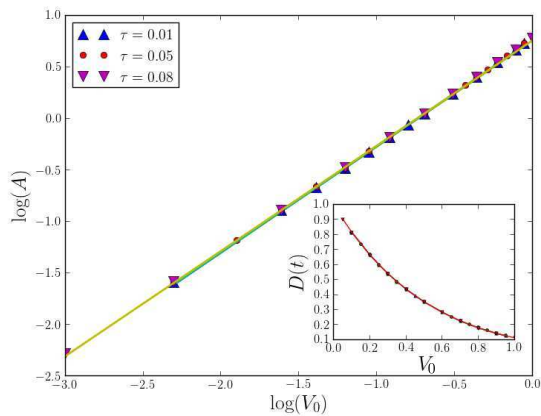}
 \end{center}
\caption{(Color online)Numerically obtained scaling (at fixed $t$): $\log~D\sim V_0$. It can be seen that the curves for different values of $\tau$ coincide with each other highlighting the weak dependence of $D(t)$ on $\tau$. Inset shows the variation of the decoherence factor of the qubit at a fixed $t$ with $V_0$.}
\label{fig_deco_scaling}
	\end{figure}

The scaling of the logarithm of the decoherence factor with $V_0$ near the QCP ($V_0=0$) is analyzed in the following way; $D(t)$ shows a Gaussian fall in time and is of the form $D(t) \sim e^{-V_0^{\lambda}t^2}$. 
We define a quantity $A(V_0) = -\log D(t)=V_0^{\lambda}t^2$ and plot $\log A(V_0)$ versus $\log V_0$ for a fixed $t$ (see Fig.~(\ref{fig_deco_scaling})). We obtain a straight line whose slope gives us $\lambda$ as $\log A(V_0)=\lambda \log V_0$ for fixed $t$. We find $\lambda=1$ which implies $A\sim V_0$. On the other hand, when $V_0$  exceed a threshold value (=$2\pi w$), one can expand the $\tan^{-1}$ term in  Eq.~(\ref{eq_decofinal}) and show that $D(t)$ is approximately independent of $V_0$. Moreover, Fig.~(\ref{fig_deco_t}) also shows that $\log D(t)$ depends very weakly on  $\tau$ for $\tau \ll 1$, which is further illustrated in Fig.~(\ref{fig_deco_scaling}) 
where we show that curves for different values of $\tau$ fall on top of each other.

\subsection{Decoherence in the sudden quench limit}\label{subsec:decosq}

The time evolution of decoherence factor can be  derived exactly from  the overlap of the initial wave function and the time-evolved version of the final wave function 
in the sudden quench ($\tau=0$) limit of
the HRZ scheme as shown below.  In this limit,  the onsite the potential is $V=0$ for $t<0$ and it is abruptly changed to $V_0+\de$
(or $V_0-\de$ depending on the initial state of the qubit) at $t=0$.
The ground state of the HCB system at $t=0$  is $\vert\psi(0)\rangle = \vert k \rangle$, which can be written with the Bogoliubov parameters as 
\beq
\vert k\rangle = \cos\theta_k^{\pm}\vert g^{\pm}(0)\rangle -\sin\theta_k^{\pm}\vert e^{\pm}(0)\rangle;
\eeq
Since the ground state and the excited state evolve in time with the corresponding eigenenergies, the wave function for $t >0$ is simply given by
\beq
\vert\psi_k^{\pm}(t)\rangle = \cos\theta^\pm_k e^{iE_k^\pm t}\vert g^{\pm}(0)\rangle - \sin\theta^\pm_k e^{-iE_k^\pm t}\vert e^{\pm}(0)\rangle.
\eeq
Expressing $\vert g^{\pm}(0)\rangle$ and $\vert e^{\pm}(0)\rangle$ in terms of $\vert k\rangle$ and $\vert k+\pi\rangle$ using the Bogoliubov transformations, we find
\beq
\begin{split}
\vert\psi_k^\pm(t)\rangle_{SQ}& = 
\vert k\rangle[\cos(E_k^\pm t)+i\sin(E_k^\pm t)\cos(2\theta_k^\pm)] \\
&+\vert k+\pi\rangle[i\sin(E_k^\pm t)\sin(2\theta_k^\pm)],
\end{split}
\label{eq_psisq}
\eeq
so that using Eq.~(\ref{eq_psisq}), we find
\beq
\begin{split}
\langle\psi_k^+(t)&\vert\psi_k^-(t)\rangle_{SQ} = \cos(E_k^+t-E_k^-t) - \\
&i\left[\sin(E_k^+t-E_k^-t)\left\{ \cos2\theta_{k} +2\delta\sin2\theta_{k}\frac{\partial \theta^-_k}{\partial \delta}\vert_{\delta=0}\right\}\right],
\end{split}
\label{eq_dksq}
\eeq
{so that we find
\begin{equation}
 \log~D(t) =  \left[\frac{-L\delta^2t^2}{2\pi}\left\{\frac{2V_0}{w}\tan^{-1}\left(\frac{2\pi w}{V_0}\right)\right\}\right].
 \label{eq_decofinalsq}
\end{equation}
Comparing Eq.~(\ref{eq_decofinal}) with Eq.~(\ref{eq_decofinalsq}),  we conclude that  the decoherence factor for the HRZ quench in the small $\tau$ limit has additional correction terms of  the order of $V_0^2\tau^2$. 
It can also be shown that for $V_0/2w>\zeta (\approx 0.75)$,  the decay constant dictating the Gaussian decay  of the  decoherence factor 
in the early time limit  is greater than the sudden quench case in comparison to the HRZ case with small $\tau$.
}


\section{Concluding Comments}\label{sec:conclusion}
 In this paper, we have studied the QPT and dynamics of  a one dimensional HCB system
 in the presence of  an onsite potential  which alternates in sign from site to site.
  We have shown that the ground state quantum fidelity shows a sharp dip at the QCP ($V=0$) indicating that the system is  in the  MI phase
 for any non-zero value of $V$. At the same time the fidelity susceptibility is also found to diverge with the system size
 in a power-law fashion dictated by the critical exponent $\nu$ (which is unity in the present case).

Subsequently, we have studied the
local von Neumann entropy density and diagonal entropy of the HCB chain in MI phase 
following the HRZ quench starting from the SF phase. The  von Neumann entropy density $s$ scales linearly with
$V_0$ for small values of $V_0$ (i.e., $V_0 < 2w$) while it becomes  independent of $V_0$ for higher values of $V_0$. 
On the other hand,  $s$ is found
to scale  quadratically  with  $\tau$ throughout.  
Interestingly,  the von Neumann entropy density  is found to be  less than its expected value of   $\log 2$ in the MI phase. This is a
consequence of the fact the system is quenched to the MI phase from the SF phase   (which is also the QCP) with $s=0$ at a finite rate which leads to  defects  in the MI phase resulting in  a surviving
supercurrent \cite{klich07} and reduced  local entropy.

We have also calculated  supercurrent and  von Neumann entropy after a FRZ quench  when the HCB chain is again brought back to the SF phase; interestingly the reduction in the supercurrent and $s$ both scale identically as ($V_o \tau)^2$ in the SF phase
emphasizing their close connection. It should also be reiterated that following the HRZ quenching there is  a surviving supercurrent as well as
 a reduction in $s$ in the MI phase. On the other hand, for the FRZ scheme  it  is the other way round; one finds  a reduction in supercurrent and a surviving  $s$ in the final SF phase. 

Finally, we have
analyzed the scaling of the decoherence factor $D(t)$ of a central qubit which is globally connected  to the
 HCB system that is driven from the initial SF phase to 
the MI phase following the HRZ quenching scheme. In the limit of a weak coupling between the qubit and
the environmental HCB system and small $\tau$, a threshold value of the magnitude of the alternating potential given by  $V_0=2\pi w$,
is found to exist. Interestingly, the 
decoherence factor grows linearly with $V_0$ when $V_0 < 2\pi w$, whereas
for $V_0 > 2 \pi w$, it turns to be independent of $V_0$. On the other hand,
 $D(t)$ is found to depend very weekly on the 
quenching parameter $\tau$. This is due to the fact that the energy spectrum  of the Hamiltonian in  the MI phase reached through
the HRZ quenching scheme depends only on $V_0$ and  not on $\tau$. 
In the sudden quench limit ($\tau \to 0$) an exact expression of  $D(t)$ is obtained.  In the case of  a finite but small
$\tau$, the decoherence factor, though qualitatively similar to the SQ case,  contains additional correction 
terms  (scaling as  $(V_0 \tau)^2$).
The more interesting observation is that we find the existence  of a threshold value of $V_0 (=1.5w)$, above which
there is a faster early time decay of the   decoherence factor of  the
HRZ case for small  $\tau$ in comparison to the sudden quenching case. Therefore, above this threshold there is  less mixing 
in the quantum state of  the qubit in the sudden quench case as compared to that of the HRZ case.

\begin{center}
{\bf Acknowledgement}
\end{center}
AD acknowledges CSIR, New Delhi for partial support through a project.


\end{document}